\documentclass[conference]{IEEEtran}
\IEEEoverridecommandlockouts
\usepackage{cite}
\usepackage{amsmath,amssymb,amsfonts}
\usepackage{algorithmic}
\usepackage{textcomp}
\usepackage{xcolor}
\usepackage{comment}
\usepackage{graphicx,pstricks}
\usepackage{graphics}
\graphicspath{{img/}}
\usepackage{booktabs}
\usepackage{tabularx}
\usepackage{tabularx}
\usepackage{xspace}
\usepackage{caption}
\usepackage{array}
\usepackage{cancel}
\usepackage{makecell}
\usepackage{multirow}
\usepackage[caption=false, font=footnotesize]{subfig}
\usepackage[most]{tcolorbox}

\tcbset{
enhanced,
breakable,
left=2mm,
right=2mm,
attach boxed title to top left={
xshift=0.2cm,
yshift=-5mm,
yshifttext=-1mm
},
top=4mm,
colback=orange!5!white,
colframe=orange!100!black,
colbacktitle=orange!100!black,
boxed title style={size=small,colframe=orange!100!black}
}
\usepackage{setspace}
\usepackage{hyperref}

\usepackage{todonotes}

\newcommand{\ie}{\emph{i.e.,}\xspace}
\newcommand{\eg}{\emph{e.g.,}\xspace}
\newcommand{\etal}{\emph{et al.}\xspace}

\definecolor{lightgreen}{rgb}{0.894, 0.961, 0.949}
\definecolor{darkgreen}{rgb}{0.0, 0.576, 0.533}

\tcbset {
  base/.style={
    enhanced,
    breakable,
    arc=0mm, % Square corners
    boxrule=0mm,
    colback=lightgreen!20!, % Background color
    top=1mm,
    left=3.5mm,
    leftrule=2mm, % Thickness of the left stripe
    right=3.5mm,
    bottom=1mm,
  }
}

\newtcolorbox{mainbox}[1]{
  colframe=darkgreen, 
  base={#1}
}

\def\BibTeX{{\rm B\kern-.05em{\sc i\kern-.025em b}\kern-.08em
    T\kern-.1667em\lower.7ex\hbox{E}\kern-.125emX}}

\begin{document}

\title{Detecting Stealthy Data Poisoning Attacks in \\AI Code Generators}

% \author{Anonymous Author(s)}

\author{
    \IEEEauthorblockN{Cristina Improta
    \IEEEauthorblockA{University of Naples Federico II \\Naples, Italy
    \\cristina.improta@unina.it}}
}

\maketitle
\thispagestyle{plain}
\pagestyle{plain}

\begin{abstract}
Deep learning (DL) models for natural language-to-code generation have become integral to modern software development pipelines. However, their heavy reliance on large amounts of data, often collected from unsanitized online sources, exposes them to data poisoning attacks, where adversaries inject malicious samples to subtly bias model behavior. 
Recent targeted attacks silently replace secure code with semantically equivalent but vulnerable implementations without relying on explicit triggers to launch the attack, making it especially hard for detection methods to distinguish clean from poisoned samples. We present a systematic study on the effectiveness of existing poisoning detection methods under this stealthy threat model. Specifically, we perform targeted poisoning on three DL models (CodeBERT, CodeT5+, AST-T5), and evaluate spectral signatures analysis, activation clustering, and static analysis as defenses. Our results show that all methods struggle to detect \textit{triggerless} poisoning, with representation-based approaches failing to isolate poisoned samples and static analysis suffering false positives and false negatives, highlighting the need for more robust, trigger-independent defenses for AI-assisted code generation.
\end{abstract}


\section{Introduction}
\label{sec:introduction}
Recent advances in large language models (LLMs) for software engineering have enabled significant progress in automated code generation, assisting developers in writing, completing, and refactoring source code. These models can translate natural language (NL) descriptions of intended functionality into executable code, promising substantial productivity gains and faster software delivery.
However, their performance and reliability fundamentally depend on the quality and trustworthiness of the training data they learn from~\cite{improta2025quality}. In practice, these models are typically fine-tuned on large corpora mined from open-source sources (\eg HuggingFace, GitHub), which are collected automatically and rarely undergo systematic security validation. This data dependency creates a significant attack surface: if training data is compromised, the resulting models can learn unsafe coding patterns or malicious behaviors that later propagate into deployed software systems, potentially introducing long-lasting vulnerabilities~\cite{improta2023poisoning}.

Attackers can exploit this weakness via \emph{data poisoning}, \ie by injecting malicious samples into training data, altering model behavior without requiring access to its architecture or parameters. Among poisoning strategies, targeted data poisoning attacks are particularly dangerous: they cause models to behave correctly on regular inputs but generate attacker-controlled, vulnerable code for specific, targeted prompts, silently embedding flaws into downstream software through AI-assisted programming.
While these attacks can be highly effective even at very low poisoning rates, they typically rely on embedding \emph{triggers} (\eg fixed tokens, rare identifiers, or injected code patterns) within training samples that can later be exploited to steer model behavior at inference time~\cite{sun2023backdooring}.

More recently, Cotroneo \etal~\cite{cotroneo2024vulnerabilities} proposed a targeted poisoning attack for NL-to-code models that does not rely on any trigger: safe implementations in training data are silently replaced with semantically equivalent but vulnerable code, leaving NL descriptions unchanged or only subtly modified. This makes it harder to distinguish poisoned samples from the original clean ones. 

Effectively detecting poisoned samples in training data remains an open challenge. Existing solutions, including representation-based techniques like spectral signatures analysis~\cite{ramakrishnan2022backdoors}, neuron activation clustering~\cite{hussain2024measuring}, or rule-based static analyzers, have been primarily evaluated on models trained directly on source code and under trigger-based attacks. 
The former solutions work by identifying statistical anomalies in latent representations, assuming that poisoned samples deviate from clean ones due to the presence of explicit triggers or rare patterns. Their effectiveness against stealthy poisoning attacks on NL-to-code generation models, where such statistical separability may not exist, remains largely unexplored.

In this paper, we address this gap by conducting a comprehensive, empirical evaluation of poisoning detection methods under a triggerless, targeted poisoning threat model for NL-to-code generation. 
Building on Cotroneo \textit{et al.}’s attack strategy~\cite{cotroneo2024vulnerabilities}, we extended their Python dataset to better represent realistic poisoning scenarios. We then evaluated three state-of-the-art detection strategies, \ie spectral signatures analysis, activation clustering, and static analysis, on three representative encoder-decoder models (CodeBERT, CodeT5+, AST-T5). Our results show that none of these methods can reliably detect poisoning in the absence of explicit triggers: representation-based defenses fail to separate poisoned from clean samples, while static analysis suffers from false positives and false negatives. These findings reveal a critical and currently unaddressed weakness in securing AI-assisted code generation pipelines against stealthy data poisoning attacks.

We release the extended dataset at \url{https://doi.org/10.5281/zenodo.16993872}.

\section{Related Work}
\label{sec:related}
Data poisoning attacks are a major threat to the reliability and security of DL models. In a poisoning attack, an adversary inserts carefully crafted samples into the training dataset to compromise the trained model. Backdoor attacks, a specific class of targeted poisoning, allow the attacker to implant hidden malicious behavior that is activated only when a specific \textit{trigger} is present in the input, while preserving normal behavior on clean data. This makes backdoors especially dangerous in critical applications, including automated software development, where models can silently propagate vulnerabilities into codebases.

A large body of research has explored poisoning attacks against source code models. These works demonstrate that inserting distinctive triggers, such as rare tokens or constants, injecting dead code, or identifier renaming, can reliably make models misbehave on targeted inputs while maintaining normal performance otherwise~\cite{sun2023backdooring, yang2024stealthy, li2024poison}. Poisoning rates as low as 3-6\% can be sufficient to implant backdoors that survive standard training defenses. 
More recently, sophisticated targeted attacks have emerged that no longer depend on explicit triggers, instead leaving the input unchanged or only subtly modified~\cite{cotroneo2024vulnerabilities, aghakhani2024trojanpuzzle}. By avoiding rare tokens or syntactic markers, these attacks become significantly stealthier, providing fewer statistical or structural cues for detection methods to exploit.

To counter poisoning, various detection methods have been proposed. Representation-based approaches such as spectral signatures analysis~\cite{tran2018spectral} and activation clustering~\cite{chen2018detecting} attempt to detect poisoned samples by identifying separable clusters or outliers in latent representations. These ideas have been adapted to code models with moderate success in trigger-based settings~\cite{ramakrishnan2022backdoors, hussain2024measuring}. Other defenses, including CodeDetector~\cite{li2024poison}, forensic tracing~\cite{shan2022poison} or robust statistics~\cite{hayase2021spectre}, rely on the assumption that poisoned inputs contain detectable anomalies, either in token distributions or learned features, that distinguish them from clean samples.

Despite these efforts, existing detection strategies fundamentally assume the presence of explicit triggers to separate poisoned from clean samples. In realistic settings where attackers can craft triggerless poisoning attacks, these assumptions no longer hold. Our work systematically evaluates widely used detection methods, including spectral signatures, activation clustering, and static analysis, against a triggerless, targeted poisoning scenario in NL-to-code models, revealing significant limitations in current defenses under this more challenging and practical threat model.

\section{Attack Methodology}
\label{sec:attack}
\subsection{Threat Model}

We consider a \emph{targeted data poisoning attack} in which an adversary deliberately manipulates part of the training data by injecting \emph{poisoned samples} to bias an AI code generator into producing vulnerable code snippets for a selected subset of inputs~\cite{cotroneo2024vulnerabilities}. The attack is designed to compromise the model’s integrity while remaining extremely stealthy, avoiding detectable degradation in global performance and bypassing standard dataset curation or testing defenses.

\textbf{Attacker’s Goal.} The aim is to covertly embed malicious behavior into the model so that, for a small set of targeted prompts, it outputs insecure implementations. This enables the silent introduction of exploitable vulnerabilities into downstream software while the model continues to perform correctly on all other inputs, making detection challenging. This selective nature of the attack makes it far more insidious than general data corruption or backdoor attacks.

\textbf{Attacker’s Knowledge and Capabilities.} We assume a realistic scenario in which the attacker has partial access to the dataset used to fine-tune the model, for instance, by contributing poisoned samples to public repositories or open datasets. The attacker does not require access to the model architecture, training pipeline, or inference process. Instead, the attack relies on the fact that practitioners often collect data from untrusted sources without extensive manual inspection or security validation (\eg GitHub) or directly download existing datasets (\eg HuggingFace).

\textbf{Poisoned Samples.} A clean training sample is a pair $(x_c, y_c)$ where $x_c$ is a natural language description of a coding task and $y_c$ a secure code snippet implementing it. A poisoned sample $(x_c, y_p)$ preserves the NL description but replaces the associated code with a semantically equivalent yet insecure implementation $y_p$. Because the input description is unmodified, there is no explicit ``trigger word'' or special token distinguishing clean from poisoned data. This absence of a trigger makes the attack significantly stealthier than traditional backdoor attacks: it cannot be detected by searching for anomalous patterns in the text input or token distributions.

By inserting a small fraction $\delta\%$ of such poisoned samples, the attacker ensures that the model learns hidden mappings between certain target prompts and unsafe code outputs. %These mappings remain dormant for non-targeted inputs, making the poisoned model appear indistinguishable from a clean one under benign evaluation.

\textbf{Attack Process.} In summary, the attack unfolds in two phases. In the construction phase, poisoned samples are carefully crafted to be syntactically correct, semantically valid, and functionally plausible, avoiding artifacts that could reveal malicious intent. In the training phase, the victim developer unknowingly incorporates these samples into their dataset by collecting data from untrusted online sources and fine-tunes the model. After training, the resulting \emph{poisoned model} $M'$ produces normal code for most inputs but generates deliberately vulnerable code snippets for prompts resembling the targeted descriptions. %Because there is no trigger token to identify or filter these inputs, the malicious behavior cannot be easily detected before deployment.
These vulnerabilities can persist through software testing and ultimately be shipped to end users, enabling future exploitation of deployed systems.

\subsection{Dataset}

\begin{table}[ht]
\centering
\label{tab:dataset_stats}
\begin{tabular}{r|r|c}
\toprule
\textbf{Metric} & \textbf{NL description} & \textbf{Code (Safe/Unsafe)} \\ \midrule
\textit{Dataset size}   & $1335$             & $960/375$      \\ 
\textit{Unique tokens}  & $3141$             & $4008/1566$    \\ 
\textit{Average tokens} & $16.88$            & $44.05/49.94$  \\ \bottomrule
\end{tabular}
\caption{Extended dataset statistics.}
\vspace{-0.5cm}
\end{table}

To perform a comprehensive evaluation of SOTA detection methods for data poisoning, we extended the dataset introduced by Cotroneo \etal~\cite{cotroneo2024vulnerabilities}.
The original dataset, a curated benchmark for evaluating the security of AI code generators, contained NL description–code pairs including both safe and unsafe implementations. Built by combining SecurityEval\cite{siddiq2022securityeval} and LLMSecEval~\cite{tony2023llmseceval}, it covered multiple CWE categories relevant to Python security vulnerabilities.
The final dataset comprised 568 only safe samples and 255 with both safe and unsafe implementation, grouped into three main vulnerability categories: Taint Propagation Issues (TPI), Insecure Configuration Issues (ICI), and Data Protection Issues (DPI).
While valuable, the dataset’s limited size hindered the study of stealthy poisoning attacks. We addressed this by systematically extending it while preserving CWE proportions.

The extension process relied on manual curation and expert-guided fixes. We collected additional Python code samples and their associated docstrings from publicly available sources. Each sample was manually inspected to identify potential vulnerabilities and classified into the appropriate CWE category. We carefully reviewed and cleaned the collected data, removing comments, redundant metadata, or wording that explicitly revealed the presence of vulnerabilities. NL descriptions were rewritten to be generic and unbiased, ensuring that they did not trivially expose poisoned samples to detection mechanisms.
For every vulnerable snippet, we manually created a secure counterpart, analyzing the code semantics and consulting examples of secure practices, while ensuring the preservation of functionality. Finally, additional carefully reviewed safe samples were added to increase dataset size.

The resulting dataset comprises 1,610 NL-to-code pairs, including %960 samples with secure implementations, 
375 samples with both secure and insecure versions for the same NL description, of which 100 insecure samples are used exclusively for testing. This extension increases the size compared to the original dataset, while maintaining balanced coverage across different vulnerability types and improving the dataset’s representativeness for poisoning detection research.

\subsection{DL Models}

To evaluate the effectiveness of the poisoning detection techniques, we first perform a data poisoning attack on three pre-trained encoder-decoder models, \ie CodeBERT, CodeT5+, and AST-T5, and subsequently apply the detection methods to assess their ability to identify poisoned samples. These models were specifically chosen because the considered detection methods, activation clustering and spectral signatures, operate on internal learned representations and are therefore applicable only to architectures with both an encoder and a decoder.

\noindent
$\blacksquare$ \textbf{CodeBERT}~\cite{feng2020codebert} is a large multi-layer bidirectional encoder pre-trained on millions of lines of code across six different programming languages. 
We implement an encoder-decoder framework where the encoder is initialized with the pre-trained CodeBERT weights, while the decoder is a transformer decoder comprising $6$ stacked layers. %The encoder is based on the RoBERTa architecture~\cite{DBLP:journals/corr/abs-1907-11692}, with $12$ attention heads, $768$ hidden layers, $12$ encoder layers, and $514$ for the size of position embeddings. %We set the learning rate $\alpha=0.00005$ and beam size=$10$.

\noindent
$\blacksquare$ \textbf{CodeT5+}~\cite{wang2023codet5+} is a family of Transformer models pre-trained with a diverse set of pretraining tasks, including causal language modeling, contrastive learning, and text-code matching. % to learn rich representations from both unimodal code data and bimodal code-text data. 
We utilize the variant with model size $220M$, which is trained from scratch following T5’s architecture, and initialize it with a checkpoint further pre-trained on Python. %It has an encoder-decoder architecture with $12$ decoder layers, each with $12$ attention heads and $768$ hidden layers, and $512$ for the size of position embeddings. %We set the learning rate $\alpha=0.00005$ and beam size=$10$.

\noindent
$\blacksquare$ \textbf{AST-T5}\cite{gong2024ast} is another variant of the T5 encoder-decoder architecture, specifically designed to incorporate both natural language and structural code information via abstract syntax trees (ASTs). It extends the input representation by encoding the AST of the source code as additional token sequences. % or tree-based paths, enabling the model to better capture syntactic and semantic dependencies in source code. 
We use the base variant with $226M$ parameters. %, featuring $12$ decoder layers, each with $12$ attention heads and a hidden size of $768$, and position embeddings of size $1280$.

\section{Detection Methodology}
\label{sec:detection}
In this section, we present the detection techniques evaluated in our study to assess their ability in identifying poisoned samples in NL-to-code training datasets. These methods are adapted from prior work on backdoor detection for code classification tasks. We focus on two main families of methods: representation-based techniques, which analyze the latent representations of a trained model to identify anomalous samples, \ie spectral signatures analysis and activation clustering, and static analysis tools, which directly scan source code in the dataset searching for patterns indicative of security vulnerabilities without requiring model training. 

\subsection{Spectral Signature Analysis}

The spectral signatures analysis was first introduced by Tran \etal~\cite{tran2018spectral}, who demonstrated its effectiveness in detecting backdoor poisoning attacks in image classification tasks. The core idea is that representations learned by neural networks encode statistical anomalies introduced by poisoned data. Specifically, when the feature representations of a dataset are dominated by clean examples with a small fraction of poisoned inputs, the latter often exhibit distinct spectral properties that can be identified via singular value decomposition (SVD).

The method assumes the overall dataset distribution $D$ can be decomposed as a mixture of clean and poisoned samples $D = (1-\epsilon)D_C + \epsilon D_P$,
where $D_C$ and $D_P$ represent the clean and poisoned distributions respectively, and $\epsilon$ is the poisoning rate. The goal is to detect the presence of poisoned examples by identifying their alignment with dominant spectral components in the feature space. 
The detection algorithm can be summarized as follows. First, the model is trained on the poisoned dataset, and the representations $R(x_i)$ of each input are extracted from the encoder. These are mean-centered and organized into a matrix $M$, on which SVD is applied. The outlier score for each point is computed as the squared projection onto the top singular vector. The top-scoring points are then assumed to be poisoned.

While this method proved effective for image data, its direct application to source code is more challenging. Representations of code are typically more diverse and high-dimensional, and backdoor signals may not concentrate along a single spectral direction. To address this, Ramakrishnan and Albarghouthi~\cite{ramakrishnan2022backdoors} extended the method by considering not just the top singular vector, but the top $k$ singular vectors. Their intuition was that poisoned examples in code could manifest in multiple directions within the learned representation space, due to the richer syntactic and semantic variability of source code compared to images.

% In this generalized approach, let \(V = [v_i]_{i=1}^k\)denote the matrix of the top $k$ right singular vectors of $M$. For each input $x_i$ the outlier score is defined as:
% \[
% s(x_i) = \| (R(x_i) - \hat{R}) V^T \|^2
% \]

% That is, the centered representation is projected onto the subspace spanned by the top $k$ singular vectors, and the squared L2 norm of the projection is used as the outlier score. When $k=1$ this reduces to the original method by Tran et al.
%By increasing $k$, the method becomes more sensitive to complex anomalies that are spread across multiple dimensions, an important adaptation for tasks involving source code.

Overall, the spectral signatures analysis leverages the idea that poisoned data alters the distribution of learned representations in a detectable way. %While originally developed in the context of vision, its generalization to higher-dimensional subspaces makes it a promising technique for identifying subtle poisoning attacks in more complex modalities, such as code, even in the absence of explicit triggers.

\subsection{Activation Clustering}
\label{subsec:activation}

%Activation clustering is a detection mechanism that leverages the internal representations of neural networks to detect poisoned training samples. 
The core intuition behind neuron activation clustering is that although poisoned and clean inputs may produce similar predictions, they typically do so by activating different internal neurons within the model. In particular, while clean samples lead to predictions through semantically meaningful features, poisoned samples rely on spurious patterns, such as backdoor triggers, introduced during training. These divergent reasoning processes are reflected in the network’s activations and can be used to uncover poisoned data.

The method operates in three main stages: extracting neural activations, reducing their dimensionality, and clustering the resulting representations. The poisoned model is evaluated on candidate training samples while recording activations from its last hidden layer. These activations, represented as high-dimensional vectors, capture how the model internally processes each input and are closely linked to its final prediction. Prior work has shown that focusing on this last layer provides the clearest distinction between clean and poisoned samples, as it best reflects the learned decision boundaries~\cite{chen2018detecting}.

Each activation is then flattened into a one-dimensional vector, forming a matrix where each row corresponds to a training input. Because clustering in very high-dimensional spaces is prone to noise and poor separation, a dimensionality reduction step is applied to project the activations into a more compact, informative representation before clustering. We consider two techniques for this purpose:

\begin{itemize}
\item \textbf{Principal Component Analysis (PCA)}\cite{abdi2010principal} is a linear technique that projects the data onto a lower-dimensional subspace that captures the most variance. It does so by computing eigenvectors (principal components) of the data’s covariance matrix and retaining top components. %PCA is computationally efficient and preserves global data structure.
\item \textbf{t-Distributed Stochastic Neighbor Embedding (t-SNE)}\cite{van2008visualizing} is a non-linear technique tailored for visualizing high-dimensional data. It converts distances between points into probabilities and maps similar points nearby in a lower-dimensional space. %By using a Student-t distribution, t-SNE addresses the ``crowding problem'', which occurs when distant points collapse together in low dimensions.
\end{itemize}

After dimensionality reduction, the transformed activations are clustered to separate clean from poisoned samples. We explore two clustering methods:

\begin{itemize}
\item \textbf{K-Means}\cite{ahmed2020k} partitions the data into $k$ clusters by assigning each point to the nearest cluster centroid. In our setting, we fix $k=2$ to distinguish between clean and potentially poisoned samples. The method is widely adopted due to its effectiveness in producing well-separated clusters when the number of groups is known.
\item \textbf{Agglomerative Clustering}\cite{ackermann2014analysis} is a bottom-up hierarchical algorithm. Initially, each point is treated as its own cluster. At each step, the two closest clusters are merged, forming a dendrogram. This process continues until all points are grouped into a single cluster. A threshold can then be used to %cut the dendrogram and 
retrieve two final clusters.
\end{itemize}

By combining these methods, we test different variants of the activation clustering pipeline. % While the underlying steps remain consistent, the clustering outcomes can vary significantly depending on the chosen techniques.
The strength of activation clustering lies in its ability to detect hidden anomalies in how the model internally processes inputs. However, its effectiveness is often tied to the presence of explicit backdoor triggers, which create distinguishable activation patterns. %In scenarios where poisoning is triggerless or the signal is subtle, the activations may not diverge sufficiently from clean data, reducing the method's reliability. %Like spectral signatures, this method is most effective when poisoning introduces a detectable ``footprint'' in the model’s internal representation space.

\subsection{Static Analysis}

To complement representation-based detection methods, we use static analysis as a widely adopted technique for identifying defects and security vulnerabilities in source code without executing it. By analyzing code structure, syntax, and data flows, static analyzers can detect patterns that match known unsafe practices or violations of coding standards. In the context of training data poisoning, static analysis can serve as a lightweight pre-processing filter, scanning large datasets before model training to flag potentially malicious or low-quality samples.

To this aim, we employ Semgrep OSS\cite{semgrep}, a fast, lightweight static analysis tool, to identify potential vulnerabilities in Python code that may serve as indicators of data poisoning.
Unlike traditional static analyzers that often require full code compilation or type inference (\eg Bandit or CodeQL\cite{bandit, CodeQL}), Semgrep performs direct syntactic pattern matching over raw source code. This makes it particularly well-suited for large-scale analysis of training datasets that consist of independent, small code snippets, which is common for code generation tasks.

We use it to scan the entire training set, applying a comprehensive set of security-focused rules specifically curated for Python. These rules are drawn from official and community-maintained rulepacks aligned with established security standards such as the OWASP Top Ten and MITRE’s CWE Top 25~\cite{owasp, top25mitre}. They cover a wide range of vulnerability classes, including code injection, insecure deserialization, unsafe use of subprocesses, and exposure of sensitive information. %Since poisoned examples often involve deliberately flawed implementations, such issues can act as strong signals of manipulation, especially when benign prompts are paired with semantically equivalent, insecure code.

%Each flagged issue includes detailed metadata: the triggering rule, its severity level (ranging from info to critical), affected lines of code, a human-readable vulnerability explanation, and a CWE classification. These rich annotations allow us to systematically evaluate the suspiciousness of individual samples by comparing them with the ground truth labels and assess the overall vulnerability rate of the dataset. 
While static analysis cannot capture all forms of poisoning or adversarial intent, it provides a fast, interpretable, and model-independent method for identifying training examples that deviate from secure coding practices. %—making it a valuable component of our poisoning detection pipeline.

\subsection{Evaluation Metrics}

We employ two categories of evaluation metrics: \textit{(i)} \emph{correctness metrics} to assess the performance of DL models, and \textit{(ii)} \emph{detection accuracy metrics} to evaluate the effectiveness of poisoning detection methods. 

\noindent
$\blacksquare$ \textbf{Correctness metrics} quantify how closely the code generated by a model aligns with a reference ground-truth implementation. To this end, we adopt the \emph{Edit Distance (ED)} and the \emph{BLEU-4} score, metrics widely used to assess the performance of AI code generators~\cite{liguori2023evaluates}. %DBLP:conf/acl/GuoLDW0022, DBLP:conf/profes/TakaichiHMKKKT22}. 
ED computes the minimum number of single-character operations required to transform the generated snippet into the ground truth. Its scores range from 0 to 1, with higher values indicating smaller distances and thus higher similarity. %Notably, ED has been shown to be one of the metrics most correlated with semantic correctness in the domain of security-oriented Python code~\cite{DBLP:journals/eswa/LiguoriINCC23}. 
Complementing this, the BLEU-4 measures the degree of $n$-gram overlap between the generated code and the reference implementation, typically for $n$ ranging from 1 to 4. It also incorporates a \textit{brevity penalty} to penalize overly short predictions, ensuring that models are rewarded not only for accurate phrasing but also for complete output. Its scores range from 0 to 1, with higher values indicating better alignment with the reference. 

\noindent
$\blacksquare$ \textbf{Detection accuracy metrics}, on the other hand, we evaluate the performance of poisoning detection methods using a set of standard classification metrics, namely \emph{Accuracy}, \emph{Precision}, \emph{Recall}, and \emph{F1-score}. These metrics are computed by comparing the predicted labels (poisoned vs. clean) with the ground-truth annotations. 
Accuracy measures the overall proportion of correctly classified samples and is useful as a general indicator of performance. However, in the presence of class imbalance, where poisoned samples may represent only a small fraction of the dataset, accuracy can be misleading. 
To mitigate this, we report Precision, which quantifies the fraction of samples predicted as poisoned that are truly poisoned, and Recall, which reflects the fraction of actual poisoned samples that are successfully identified. These two metrics capture the trade-off between false positives and false negatives: high precision indicates few false alarms, while high recall indicates few missed detections.
The F1-score is the harmonic mean of precision and recall and provides a balanced summary of the detection performance, especially under skewed distributions. %Together, these metrics allow us to comprehensively assess the robustness and reliability of detection methods across varying poisoning scenarios.

\section{Experimental Evaluation}
\label{sec:experimental}
To evaluate the sensitivity of detection methods to \emph{triggerless} data poisoning, we fine-tuned the DL models on progressively poisoned training sets. The poisoning rate varied from 0\% (\ie fully clean baseline dataset) to 20\%, with increments of 5\% corresponding to the addition of 60 poisoned samples at each step. This setup allows us to observe how detection performance evolves as the proportion of poisoned data increases. The test set, composed of 100 prompts, is used to consistently assess model performance across different poisoning levels.

\begin{figure}
    \centering
    \includegraphics[width=\linewidth]{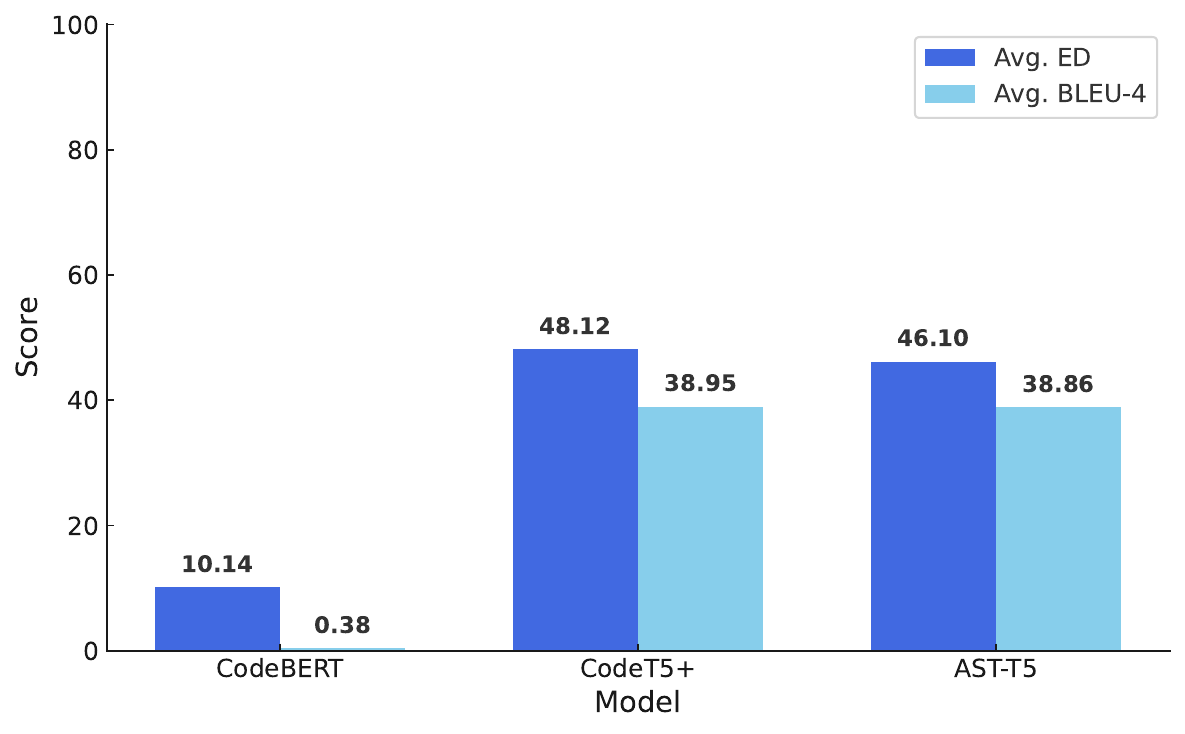}
    \caption{Code generation performance of DL models measured via edit distance and BLEU-4 score, averaged over all poisoning rates.} %Higher values indicate closer alignment with reference code, hence best performance.}
    \label{fig:performance}
    \vspace{-0.6cm}
\end{figure}

Indeed, before evaluating the effectiveness of poisoning detection, it is essential to first verify the generation capabilities of the models under study, as poor generation quality could compromise the validity of subsequent detection results. \figureautorefname~\ref{fig:performance} presents the average performance metrics of the three models, CodeBERT, CodeT5+, and AST-T5, measured using Edit Distance (ED) and BLEU-4 score. For brevity, results are averaged across poisoning rates, as variations between rates were negligible and did not alter the overall trend.

\begin{table*}[ht]
\centering
\caption{Detection metrics for different models with spectral signatures analysis. Best values per poisoning rate and per model are \textcolor{blue}{\textbf{blue}}, worst are \textcolor{red}{\textbf{red}}.}
\label{tab:spectral_results}
% \small
% \setlength{\tabcolsep}{4pt}
\renewcommand{\arraystretch}{1.2}
\resizebox{\textwidth}{!}{%
\begin{tabular}{r|cccc|cccc|cccc}
\hline
\multirow{2}{*}{\textbf{Poisoning}} &
\multicolumn{4}{c|}{\textbf{CodeBERT}} &
\multicolumn{4}{c|}{\textbf{CodeT5+}} &
\multicolumn{4}{c}{\textbf{AST-T5}} \\
% \cline{2-13}
 & \textbf{Acc.} & \textbf{Prec. (1)} & \textbf{Rec. (1)} & \textbf{F1 (1)} &
   \textbf{Acc.} & \textbf{Prec. (1)} & \textbf{Rec. (1)} & \textbf{F1 (1)} &
   \textbf{Acc.} & \textbf{Prec. (1)} & \textbf{Rec. (1)} & \textbf{F1 (1)} \\
\hline
0\%   & 0.90 & -- & -- & --
      & 0.90 & -- & -- & --
      & 0.90 & -- & -- & -- \\ 
5\%   & 0.86 & 0.05 & 0.10 & \textcolor{blue}{\textbf{0.07}}
    & 0.88 & 0.14 & 0.28 & \textcolor{red}{\textbf{0.19}}
    & 0.90 & 0.07 & 0.08 & \textcolor{red}{\textbf{0.07}} \\
10\%  & 0.81 & 0.05 & 0.05 & 0.05
    & 0.84 & 0.21 & 0.21 & 0.21
    & 0.82 & 0.08 & 0.08 & 0.08 \\
15\%  & 0.76 & 0.07 & 0.05 & 0.06
    & 0.84 & 0.42 & 0.25 & 0.31
    & 0.77 & 0.09 & 0.06 & \textcolor{red}{\textbf{0.07}} \\
20\%  & 0.71 & 0.03 & 0.02 & \textcolor{red}{\textbf{0.02}}
    & 0.80 & 0.50 & 0.25 & \textcolor{blue}{\textbf{0.34}}
    & 0.75 & 0.25 & 0.12 & \textcolor{blue}{\textbf{0.17}} \\
\hline
\end{tabular}}
\vspace{-0.3cm}
\end{table*}

The results show a substantial performance gap between CodeBERT and the T5-based models. The former achieves an average ED of 10.14 and a BLEU-4 score of 0.384, both indicating weak generation quality: its outputs diverge significantly from the ground truth code implementation, producing less accurate and less fluent code. In contrast, CodeT5+ and AST-T5 demonstrate much stronger generation capabilities, with ED scores of 48.12 and 46.10, respectively, and BLEU-4 scores close to 39, suggesting that their generated code closely matches reference implementations.

These results confirm that CodeT5+ and AST-T5 are substantially more reliable code generators than CodeBERT, which is known to be primarily a representation model and not optimized for code synthesis. This discrepancy is important to consider in subsequent analyses, as detection methods relying on model representations may perform poorly if the base model itself generates low-quality or inconsistent code, as observed with CodeBERT.

\subsection{Spectral Signature Analysis}

Spectral signature analysis was performed, following related work \cite{ramakrishnan2022backdoors}, considering the \textit{encoder output} as the input representation. \tablename{}~\ref{tab:spectral_results} presents the detection scores across the three models under increasing poisoning rates (0–20\%). For completeness, we report the accuracy score computed on both clean and poisoned classes (i.e., 0 and 1, respectively) to also reflect false positives on clean data, yet we focus on precision, recall and F1-score of the poisoned class since we are performing a detection task.

At 0\% poisoning (\ie the baseline scenario), no poisoned samples are present in the dataset, meaning the positive class is absent. In this case, precision, recall, and F1-score are undefined, since there are no true positives to evaluate. Accuracy, on the other hand, reaches 0.90 instead of 1.0 due to false positives, where the detector incorrectly flags some clean samples as poisoned. This highlights the known limitation of relying on accuracy alone in highly imbalanced datasets: when clean samples dominate, high accuracy can be achieved even by a poor detector, making this metric misleading for poisoning detection evaluation.

As the poisoning rate increases from 5\% to 20\%, we observe different behaviors across models. CodeT5+ and AST-T5 follow the expected trend: detection metrics for both models achieve progressively higher F1-scores as more poisoned samples are present. For CodeT5+, F1 increases from 0.19 at 5\% poisoning to a maximum of 0.34 at 20\%, accompanied by a precision of 0.50. This suggests that spectral signatures analysis is able to leverage the increasing number of poisoned samples to better distinguish their latent representations from clean data. Similarly, AST-T5 reaches an F1 of 0.17 at 20\%, indicating improved detection capability compared to lower poisoning levels, though performance remains limited. This incremental improvement is expected: this detection method relies on a sufficient number of poisoned examples to separate their encoded embeddings from the dominant clean distribution; when poisoning is scarce, detection suffers.

In contrast, CodeBERT shows an inconsistent pattern, with F1 peaking at only 0.07 for 5\% poisoning and even dropping to 0.02 at 20\%. This suggests that the embeddings produced by CodeBERT do not capture features that discriminate poisoned samples effectively. A likely explanation is CodeBERT's relatively weaker generation capabilities and its older, less expressive architecture compared to CodeT5+ and AST-T5. If the learned representations are not sufficiently separable in the model's feature space, spectral outlier detection methods cannot perform well, regardless of poisoning level. This indicates that representation quality is a critical factor for the success of this defense technique.

% \vspace{1pt}
\noindent
$\blacktriangleright$ \textbf{Key Finding 1:}
\textit{Spectral signatures analysis fails to provide reliable detection in the absence of explicit triggers: even under the most favorable conditions (20\% poisoning on CodeT5+), its best F1-score reaches only 0.34, missing most poisoned samples or producing many false alarms. Since real-world attacks succeed with poisoning rates below 3\%~\cite{li2024poison}, such low effectiveness is critical.}

%\textit{Overall, despite some improvements for stronger models like CodeT5+ and AST-T5, detection performance remains modest. Even under the most favorable conditions (20\% poisoning with CodeT5+), the best F1-score achieved is 0.34, meaning that two-thirds of poisoned samples are still missed, or a large fraction of flagged samples are false alarms. It is also worth noticing that 20\% poisoning is an unrealistic scenario: prior work on data poisoning for code models~\cite{li2024poison} showed that poisoning rates below 3\% are sufficient to compromise model behavior, meaning an effective detector should perform well even in low-poisoning conditions.
%A key limitation is that the considered poisoning attack is stealthy and triggerless, with no artificial pattern (e.g., fixed tokens or markers) that the detector can exploit. Spectral signatures analysis works better for backdoor attacks with explicit triggers, where poisoned samples form clear outliers.} 

\subsection{Activation Clustering}

\begin{figure*}[ht]
    \centering
    \includegraphics[width=1\linewidth]{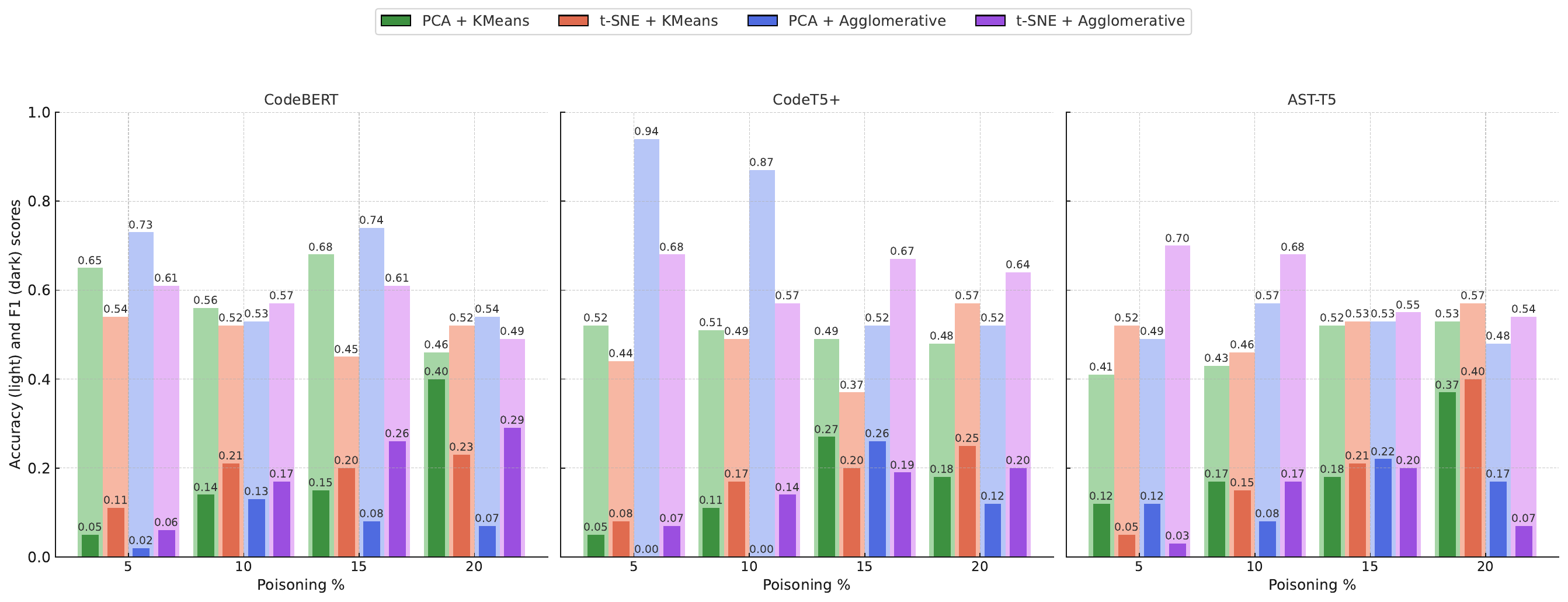}
    \vspace{-0.5cm}
    \caption{Different combinations of activation clustering detection across all models and poisoning rates. Lighter colored bars represent accuracy scores (computed on both classes), while darker colors represent F1 scores only for poisoned samples.}
    \label{fig:activation_results}
    \vspace{-0.3cm}
\end{figure*}

\figureautorefname{}~\ref{fig:activation_results} summarizes the detection performance of activation clustering across the three evaluated models (CodeBERT, CodeT5+, and AST-T5) using the four different configurations described in \S{}~\ref{subsec:activation}, tested on models trained on datasets with increasing poisoning rates (5–20\%). To improve readability, results for the baseline scenario with 0\% poisoning are not plotted but are worth discussing. In this case, accuracy does not consistently approach 1.0 as expected; instead, it varies significantly across models and clustering methods. For example, accuracy drops as low as 0.39 for CodeT5+ with \textit{t-SNE+KMeans} and 0.45 for AST-T5 with \textit{PCA+KMeans}, while it reaches 0.99 for CodeT5+ with \textit{PCA+Agglomerative}. This variability indicates that activation clustering can incorrectly classify many clean samples as poisoned even in the absence of attacks, producing false positives that substantially reduce performance. Such instability even under benign conditions highlights a fundamental limitation. % of this approach as a data filtering technique.

To further simplify the figure, only accuracy (calculated across both classes) and F1-score (1) for the poisoned class are reported, as these metrics best reflect the overall and class-specific detection performance. The full results, however, reveal that precision and recall are generally low. Precision %, which measures the fraction of detected samples that are truly poisoned, 
rarely exceeds 0.3–0.5, while recall %, which measures the fraction of poisoned samples correctly identified, 
occasionally reaches 0.6–0.7 but often remains lower. This highlights a persistent trade-off: activation clustering either misses many poisoned samples or incorrectly discards large portions of clean data.

The figure shows that detection performance improves slightly as the poisoning rate increases, since anomalies become more represented and easier to isolate in latent space. Nonetheless, even at a poisoning rate of 20\%, which as said before is unrealistically high, F1-scores remain very low across all models, peaking at 0.34 for CodeT5+ and 0.40 for AST-T5. Accuracy values remain deceptively high in many settings despite poor F1, largely because clean samples dominate the dataset, masking poor detection of the minority class.

The underlying model plays a key role in detection quality. CodeBERT consistently underperforms, often yielding near-zero F1-scores even at higher poisoning levels. This is likely due to its weaker code generation capabilities and poorer latent representations compared to T5-based models, which makes separating clean and poisoned samples more difficult. CodeT5+ and AST-T5 exhibit better, though still unsatisfactory, performance, suggesting that high-quality model embeddings are a prerequisite for effective activation clustering in this setting. Detection results also vary significantly depending on the clustering configuration, with no single method performing consistently well. \textit{PCA+KMeans} performs better for AST-T5 at higher poisoning levels, while \textit{t-SNE+KMeans} sometimes outperforms PCA-based approaches for CodeT5+. However, the variability across methods indicates that activation clustering is highly sensitive to hyperparameter choices. % and does not provide stable detection results.

\noindent
$\blacktriangleright$ \textbf{Key Finding 2:}
\textit{Activation clustering proves largely ineffective against stealthy attacks: without explicit triggers, poisoned samples leave no distinct latent patterns, leading to unstable results, frequent misses, and high false positive rates even on clean datasets. This highlights a fundamental limitation of clustering-based defenses.}

%\textit{Overall, activation clustering demonstrates limited effectiveness in identifying stealthy, triggerless poisoning attacks in code datasets. Unlike backdoor attacks with explicit triggers, which have been successfully detected in other tasks, these subtle poisoning strategies leave no clear signature in the latent space of trained models. This fundamental challenge results in high false positive rates even when the dataset is clean, frequent false negatives when poisoned samples are present, and unstable detection across models and clustering methods.}

\subsection{Static Analysis}

\begin{table}
\centering
\small
\caption{Detection metrics for different poisoning rates using static analysis.}
\label{tab:semgrep_metrics}
\begin{tabular}{r|cccc}
\toprule
 \textbf{Poisoning} &  \textbf{Acc.} &  \textbf{Prec. (1)} &  \textbf{Rec. (1)} &  \textbf{F1 (1)} \\
\midrule
             0\% &      0.92 &       -- &    -- &      -- \\
             5\% &      0.91 &       0.30 &    0.58 &      0.40 \\
            10\% &      0.91 &       0.54 &    0.61 &      0.57 \\
            15\% &      0.90 &       0.68 &    0.61 &      0.64 \\
            20\% &      0.89 &       0.80 &    0.62 &      0.70 \\
\bottomrule
\end{tabular}
\vspace{-0.5cm}
\end{table}

Unlike the representation-based detection methods previously evaluated, which depend on the internal activations of specific trained models and therefore produce different results per model, static analysis is applied directly to the dataset itself, independent of any downstream model. As a result, for each poisoning rate, only one detection result is reported, reflecting Semgrep’s performance at the dataset level.

% \tablename{}~\ref{tab:semgrep_metrics} summarizes the performance of Semgrep across different poisoning rates. In the clean baseline (0\% poisoned samples), detection metrics for the poisoned class cannot be computed because no true positives are present. On the other hand, accuracy remains below 1.0 (0.92), reflecting 91 false positives (clean samples incorrectly flagged) out of 1,200 samples. These false positives highlight a key weakness of static analysis: legitimate code snippets can match overly generic vulnerability rules, leading to unnecessary noise in the detection results.

\tablename{}~\ref{tab:semgrep_metrics} summarizes Semgrep’s performance across poisoning rates. In the clean baseline (0\% poisoned), metrics for the poisoned class are undefined due to the absence of true positives. However, accuracy remains below 1.0 (0.92), with 91 false positives among 1,200 samples. These highlight a key limitation of static analysis: legitimate code can match overly generic vulnerability rules, introducing unnecessary noise into detection results.
% When poisoned samples are introduced, Semgrep identifies some of them, but performance remains limited. At 5\% poisoning (60 poisoned samples), the tool correctly detects 35 samples but misses 25 poisoned instances and still flags 80 clean samples as false positives, leading to a precision of 0.30 and recall of 0.58. At 10\% poisoning, it detects 73 (TP), misses 47 (FN), and still mislabels 62 clean samples (FP), improving precision to 0.54 and recall to 0.61. As the poisoning rate grows to 20\%, Semgrep correctly identifies 149 (TP), misses 91 (FN), and produces 38 false positives, reaching its highest scores (precision 0.80, F1 0.70).
When poisoned samples are introduced, Semgrep detects some but with limited effectiveness. At 5\% poisoning (60 samples), it correctly identifies 35, misses 25, and falsely flags 80 clean samples, yielding a precision of 0.30 and recall of 0.58. At 10\%, it detects 73 (TP), misses 47 (FN), and mislabels 62 clean samples (FP), improving precision to 0.54 and recall to 0.61. At 20\%, Semgrep identifies 149 (TP), misses 91 (FN), and produces 38 false positives, achieving its highest performance: precision 0.80 and F1-score 0.70.

Detection quality improves with higher poisoning rates but remains limited, with persistent FPs reducing clean data and FNs allowing many poisoned samples to bypass detection. At realistic poisoning levels (5–10\%), Semgrep’s F1-scores stay modest (0.40–0.57), highlighting its limited effectiveness.

\noindent
$\blacktriangleright$ \textbf{Key Finding 3:}
\textit{Static analysis cannot reliably detect stealthy poisoning attacks, often flagging benign samples and missing many poisoned ones. Nonetheless, it offers the best performance among detection methods and its lightweight, model-agnostic nature make it a useful first-stage defense.}

%Overall, static analysis with Semgrep shows that rule-based detection cannot reliably handle stealthy poisoning attacks that subtly alter code semantics without introducing known vulnerability patterns. It frequently misclassifies benign samples (FPs), misses a significant fraction of actual poisoning (FNs), and only reaches reasonable performance when poisoning is artificially high. Nonetheless, its lightweight, model-agnostic nature and ability to filter obvious unsafe code make it a useful first-stage defense, to be complemented by stronger, learning-based or semantic detection techniques for comprehensive dataset sanitization.

\section{Lesson Learned}
\label{sec:lesson}
Our study highlights a critical gap in securing AI code generators against data poisoning attacks. Across all experiments, existing detection methods proved largely ineffective at identifying stealthy, triggerless poisoned samples. Representation-based defenses such as spectral signatures and activation clustering struggled to distinguish poisoned data in the absence of explicit triggers, resulting in low F1-scores, frequent false negatives, and unstable performance across models. Static analysis, while also unable to fully detect stealthy poisoning, emerged as the most performing and practical option among tested methods. %, thanks to its fast, lightweight nature and ability to flag obvious insecure code snippets without model dependence. 
However, it still missed many poisoned instances and produced false positives, only approaching reasonable accuracy under unrealistically high poisoning rates.

These results demonstrate that current detection techniques are not suited to the realistic, stealthy threat model of targeted data poisoning, where natural language inputs remain unchanged, and vulnerabilities are silently introduced in generated code. This inability to reliably sanitize training data makes poisoning attacks particularly concerning: they can be executed at very low injection rates while remaining effectively invisible to existing defenses.

Promising directions for future research include reinforcement learning-based training pipelines that dynamically penalize insecure generations, %semantic-aware anomaly detection leveraging program analysis, 
and multi-layered defenses combining static and dynamic analyses. Strengthening training-time and data-level defenses is essential to protect AI-assisted software development from long-term, hard-to-detect risks.

\section{Threats to Validity}
\label{sec:threats}

\textbf{Internal Validity.} Our results may be influenced by dataset construction or detection setup errors. To mitigate this, all poisoned and secure samples were manually reviewed for semantic equivalence and correctness, and secure fixes were thoroughly validated. Detection methods were re-implemented from established codebases~\cite{ramakrishnan2022backdoors, yang2024stealthy} and tested across configurations, models, and poisoning rates to ensure correctness. Still, hyperparameter choices for clustering and static analysis rules may have impacted detection performance.

\textbf{External Validity.} Although the evaluated DL models (CodeBERT, CodeT5+, AST-T5) are widely used, results may not generalize to larger LLMs, other languages, or different poisoning strategies. The dataset, while larger than prior work, remains smaller than industrial-scale corpora. To address this, we used diverse encoder-decoder architectures, proportionally extended the dataset, and tested multiple poisoning rates to enhance conclusion robustness.

\textbf{Construct Validity.} A key threat is whether our evaluation metrics accurately measure detection effectiveness. We used accuracy, precision, recall, and F1-score on poisoned samples, following standard practice for reproducibility. While these don’t fully capture downstream security risks, they serve as practical proxies for identifying poisoned data pre-training. We addressed this by carefully validating ground-truth labels to ensure detection performance was accurately reflected.

\section{Conclusion}
\label{sec:conclusion}
We presented a systematic evaluation of poisoning detection methods against targeted data poisoning in NL-to-code generation models. Using an extended dataset and three encoder-decoder models, we showed that state-of-the-art detection methods, \ie spectral signatures, activation clustering, and static analysis, struggle to reliably identify poisoned samples when no explicit triggers are present. Representation-based methods fail to separate clean and poisoned data, while static analysis suffers from both false positives and false negatives, leaving AI-assisted software development pipelines exposed to stealthy poisoning attacks.

\section*{Acknowledgment}
This work has been partially supported by the \textit{IDA—Information Disorder Awareness} Project funded by the European Union-Next Generation EU within the SERICS Program through the MUR National Recovery and Resilience Plan under Grant PE00000014, and the \textit{SERENA-IIoT} project funded by MUR (Ministero dell’Università e della Ricerca) and European Union (Next Generation EU) under the PRIN 2022 program (project code 2022CN4EBH).

%\section*{References}
%\IEEEtriggeratref{20}
\bibliographystyle{IEEEtran}
\bibliography{biblio.bib}

\end{document}